\documentclass{elsarticle}
\usepackage{hyperref}

\journal{JASTP}

\biboptions{authoryear}

\begin{document}

\begin{frontmatter}

\title{Comment on the paper by Popova et al. ``On a role of quadruple component of magnetic field in defining solar activity in grand cycles''}

\author{Ilya G. Usoskin}
\address{Space Climate Research Unit and Sodanky\"a Geophysical Observatory, 90014 Oulu, Finland}

\begin{abstract}
The paper by Popova et al. presents an oversimplified mathematical model of solar activity with a claim of
 predicting/postdicting it for several millennia ahead/backwards.
The work contains several flaws devaluating the results:
(1) the method is unreliable from the point of view of signal processing (it is impossible to make harmonic
 predictions for thousands of years based on only 35 years of data) and lacks quality control;
(2) the result of post-diction apparently contradicts the observational data.
(3) theoretical speculations make little sense;
To summarize, a multi-harmonic mathematical model, hardly related to full solar dynamo theory, is presented,
 which is not applicable to realistic solar conditions because of the significant chaotic/stochastic intrinsic
 component and strong non-stationarity of solar activity.
The obtained result is apparently inconsistent with the data in the past and thus cannot be trusted for the future predictions.
\end{abstract}


\end{frontmatter}


\section{Introduction}

I was invited by a Guest Editor of the Topical Issue \textit{``Future solar activity''} of JASTP journal to
 review the paper by \citet[][denoted as P17 henceforth]{popova17}.
Unfortunately, because of an unexpected technical problem with the publisher's online system, my review was lost during the manuscript
 processing and was not formally accounted for by the Editors when evaluating the P17 paper.
However, it appears important to inform the scientific community about this review and, specifically, about scientific problems
 related to the P17 paper.
This small Comment is written on the basis of the lost review and summarizes important flaws in the analysis method and results,
 published by P17.

\section{The method}

P17 aims to predict solar activity for 3000 year.
The prediction method is based on a simple three-harmonic model of solar activity (two dipole and one quadruple components).
The dipole components are periodic with frequencies being close to each other (21.41 and 22.62 years),
 which leads to a beating frequency of about 350--400 years.
These dipole components were ``defined'' elsewhere \citep{zharkova15} from a 35-year long set of solar data.
However, as known from data processing, frequencies cannot be defined with the necessary precision from this dataset.
For example, in order to separate, in a statistically significant way, these two frequencies, one needs about 400 years of data.
Therefore, the beating period of $\approx$400 years can not be accurately defined from such a short dataset and is a pure artefact,
 which cannot be statistically defined from the available data.
The quadruple component is introduced as a purely ad-hoc sine wave with the period chosen to obtain the third beating
 period of around 100 years.
Thus effectively, the authors of P17 represent the long-term solar activity by a multi-harmonic oscillator.
This approach would work only for a precisely known and purely stationary series.
However, this is clearly not a case for solar activity which contains an essential intrinsic chaotic/stochastic component
 \citep[e.g.,][]{kremliovsky95,petrovay10,usoskin_LR_17}.
Similar attempts to model solar variability by a multi-harmonic (also nonlinear) oscillator have been preformed since the 1950s \citep[e.g.,][]{schove55}
 but failed.
Anyway, the authors do not present any analysis of the stability and robustness of the method and provide no clue on the range of its validity.
The choice of the main beating frequencies is ungrounded and imprecise.
For example, the Gleissberg cycle is not a single 100-yr mode but rather a wide-band variability with typically two sub-modes, 70-90 years and 120-150 years
 \citep[e.g.,][]{ogurtsov02,vecchio17}.
The claimed $\approx 400$-year cycle is not pronounced in solar activity.
Instead, the very well-defined Suess/de Vries cycle of $\approx 210$-year periodicity is not present here.
It is also unclear why the authors ``limit'' themselves to the period of 1200--3000 AD?
If their method worked, they could equally ``well'' predict solar activity for tens or thousands or millions of years
 ahead/backwards, as based on an implicit assumption of the full stationarity and perfect harmonicity of the series.

\section{Validation of the results}

\begin{figure}[t]
\centering \resizebox{\textwidth}{!}{\includegraphics{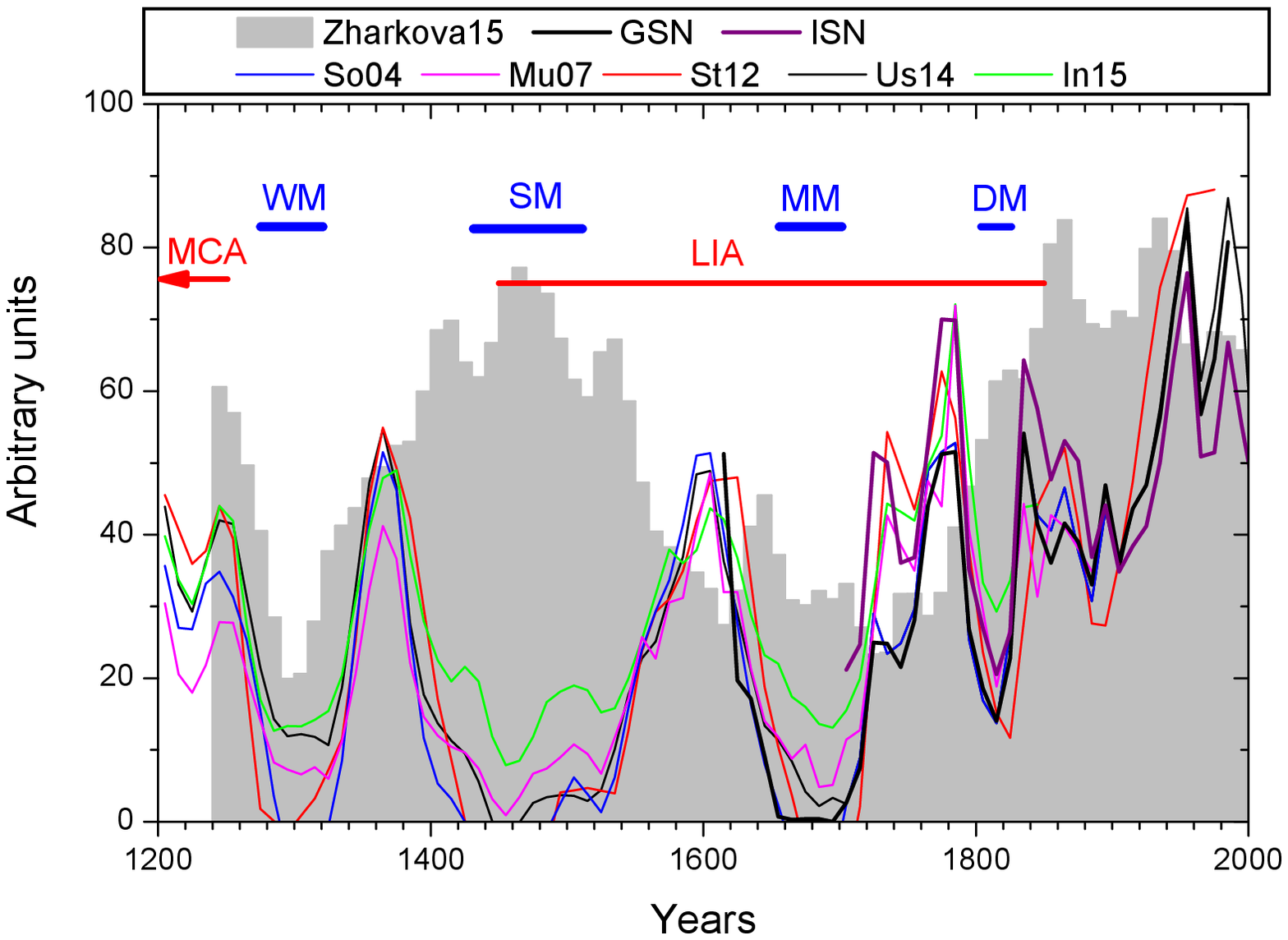}}
\caption{Comparison of the decadal variability of solar activity as reconstructed by P17 (grey shaded area) and
 other direct/indirect data: groups sunspot number GSN \citep{hoyt98}; International sunspot number v2.0 \citep{clette14};
 $^{14}$C-based reconstruction Mu16 \citep{muscheler16}; $^{10}$Be-and $^{14}$C-based reconstruction St12 \citep{steinhilber12};
 $^{14}$C-based Us14 \citep{usoskin_AAL_14}; $^{10}$Be-based one In15 \citep{inceoglu15}.
 Periods of the Wolf, Sp\"orer, Maunder and Dalton minima are indicated by blue lines.
}
\label{Fig:zh}
\end{figure}

The result of the P17 paper factually voids the prediction by \citet{zharkova15} as appears obvious from Figs. 2 and 3,
 and the authors should have said clearly that their earlier results were not correct.
However, even the new result disagrees with the available data for the last centuries.
While the authors did not show a direct comparison between their results and other direct/indirect data on solar activity,
 I do it here in Figure~\ref{Fig:zh} for decadally averaged data (modulus of the final prediction shown in Figure~3 of P17) versus
 different other reconstructions, based on sunspot counts/drawing and cosmogenic isotopes.
While Dalton and partly Maunder minima are somehow reproduced by the P17 model, the Sp\"orer minimum in the 15-16th century is totally
 missed by the method, which instead forecasts a very high activity comparable to that in the 20th century.
In fact, any ``noisy'' time series with approximately the correct autocorrelation can match some of the variations
 purely by chance.
However, the P17 authors have not demonstrated that the agreement between their model and the sunspot number is better than a chance,
 as discussed below.
The Sp\"orer minimum was one of the deepest and longest grand minima of solar activity (bigger than the Maunder minimum), and its existence
 is beyond any doubts as follows from numerous independent results based on cosmogenic nuclides $^{10}$Be and $^{14}$C
 \citep[e.g.,][]{beer12,steinhilber12,usoskin_AAL_14,inceoglu15}.
We are aware of an attempt of the P17 author to ``demolish'' the Sp\"orer minimum \citep{zharkova17}, but it is not yet published in a refereed journal
 and also contains serious flaws to be addressed beyond this Comment upon publication of that work.
Thus, we have no reason to believe in the non-existence of the Sp\"orer minimum.

The failure of the method to reproduce a major grand minimum of solar activity five hundred years ago invalidates any predictive capability of the model.
Moreover, it is not only the Sp\"orer minimum which is not reproduced.
The overall P17 result shows no statistically significant correlation with other series.
For example, the Pearson correlation coefficient (no time shift) between the P17\footnote{Modulus of the magnetic field
 amplitude scanned from Figure 3a of P17.} and Mu16, shown in Figure~\ref{Fig:zh}, curves for the period 1200--1900 is 0.16
 which is an insignificant \footnote{Significance of the correlation is estimated using the non-parametric random phase method
 \citep{ebisuzaki97,usoskin_JASTP_06}, since the standard formulas are not applicable because of the
 high level of autocorrelation in the compared series.} correlation ($p\approx 0.2$).
The correlation between P17 and ISN curves for the period 1700--2000 (viz. excluding the Sp\"orer minimum) is also insignificant (0.33, $p=0.16$).
For comparison, the correlation between Mu17 and ISN series for the period 1700--1900 is highly significant (0.64, $p=0.007$).

Thus, the method is unable to reproduce the observed solar variability for the last centuries, which makes any prediction from
 this model completely unreliable.

\section{Theoretical speculations}

Theoretical speculations by P17 make little sense and are hardly relevant.
The ad-hoc introduced quadruple component is vague.
The authors wanted to add a third harmonic component to their model, but it is ungrounded why it should be a quadruple mode.
The authors state that they are unable to find this mode in the real solar magnetic data and introduce it just out of the blue.
Moreover, substituting the full dynamo equations with equations for ``selected modes'' is a dangerous exercise which can easily
 lead to a spurious result.
The authors have not provided solid arguments that such a substitution is representative for solar activity on long-term scale.
While the two main components are at least based on data (though in a non-rigorous manner), the third component has no clear meaning.

\section{Summary}

Accordingly, as discussed above, the paper P17 contains several flaws which make the prediction of solar activity for the next thousands years
 unreliable.
\begin{itemize}
\item
The method of P17 is based on an oversimplified and unreliable ad-hoc multi-harmonic representation of solar activity, and lacks quality control.
In particular, the background solar dataset (35 years) does not allow determination of periodicities with sufficient accuracy to justify the
 beating period of 400 years.
It is therefore impossible to make harmonic predictions for thousands of years based on only 35 years of data.
\item
 The result of the post-diction contradicts the observational data of past solar activity.
 In particular, it fails to reproduce the greatest grand minimum of solar activity, Sp\"orer minimum,
  and also does not correlate with the known variability of solar activity in a statistically significant manner.
\item
Theoretical speculations make little sense.
In particular, the third quadruple component of the model is introduced purely ad-hoc with the purpose of obtaining a beating period of 100 years.
\end{itemize}

To summarize, a multi-harmonic mathematical model, hardly related to full solar dynamo theory, is presented,
 which is not applicable to realistic solar conditions because of the significant chaotic/stochastic intrinsic
 component and strong non-stationarity of solar activity.
The obtained result is apparently inconsistent with the data in the past and thus cannot be trusted for the future predictions.

\section*{Acknowledgement}
\noindent
This work was made in the framework of ReSoLVE Centre of Excellence (Academy of Finland, project no. 272157).

\section*{References}


\begin{thebibliography}{18}
\expandafter\ifx\csname natexlab\endcsname\relax\def\natexlab#1{#1}\fi
\providecommand{\url}[1]{\texttt{#1}}
\providecommand{\href}[2]{#2}
\providecommand{\path}[1]{#1}
\providecommand{\DOIprefix}{doi:}
\providecommand{\ArXivprefix}{arXiv:}
\providecommand{\URLprefix}{URL: }
\providecommand{\Pubmedprefix}{pmid:}
\providecommand{\doi}[1]{\href{http://dx.doi.org/#1}{\path{#1}}}
\providecommand{\Pubmed}[1]{\href{pmid:#1}{\path{#1}}}
\providecommand{\bibinfo}[2]{#2}
\ifx\xfnm\relax \def\xfnm[#1]{\unskip,\space#1}\fi
\bibitem[{Beer et~al.(2012)Beer, McCracken \& von Steiger}]{beer12}
\bibinfo{author}{Beer, J.}, \bibinfo{author}{McCracken, K.}, \&
  \bibinfo{author}{von Steiger, R.} (\bibinfo{year}{2012}).
\newblock {\it \bibinfo{title}{Cosmogenic Radionuclides: Theory and
  Applications in the Terrestrial and Space Environments}\/}.
\newblock \bibinfo{address}{Berlin}: \bibinfo{publisher}{Springer}.
\bibitem[{Clette et~al.(2014)Clette, Svalgaard, Vaquero \& Cliver}]{clette14}
\bibinfo{author}{Clette, F.}, \bibinfo{author}{Svalgaard, L.},
  \bibinfo{author}{Vaquero, J.}, \& \bibinfo{author}{Cliver, E.}
  (\bibinfo{year}{2014}).
\newblock \bibinfo{title}{Revisiting the sunspot number: A 400-year perspective
  on the solar cycle}.
\newblock {\it \bibinfo{journal}{Space Sci. Rev.}\/},  {\it
  \bibinfo{volume}{186}\/}, \bibinfo{pages}{35}.
  \DOIprefix\doi{10.1007/s11214-014-0074-2}.
\bibitem[{{Ebisuzaki}(1997)}]{ebisuzaki97}
\bibinfo{author}{{Ebisuzaki}, W.} (\bibinfo{year}{1997}).
\newblock \bibinfo{title}{{A Method to Estimate the Statistical Significance of
  a Correlation When the Data Are Serially Correlated.}}
\newblock {\it \bibinfo{journal}{J. Climate}\/},  {\it \bibinfo{volume}{10}\/},
  \bibinfo{pages}{2147--2153}.
  \DOIprefix\doi{10.1175/1520-0442(1997)010<2147:AMTETS>2.0.CO;2}.
\bibitem[{Hoyt \& Schatten(1998)}]{hoyt98}
\bibinfo{author}{Hoyt, D.~V.}, \& \bibinfo{author}{Schatten, K.~H.}
  (\bibinfo{year}{1998}).
\newblock \bibinfo{title}{Group sunspot numbers: A new solar activity
  reconstruction}.
\newblock {\it \bibinfo{journal}{Solar Phys.}\/},  {\it
  \bibinfo{volume}{179}\/}, \bibinfo{pages}{189--219}.
  \DOIprefix\doi{10.1023/A:1005007527816}.
\bibitem[{Inceoglu et~al.(2015)Inceoglu, Simoniello, Knudsen, Karoff, Olsen,
  Turck-Chi\'eze \& Jacobsen}]{inceoglu15}
\bibinfo{author}{Inceoglu, F.}, \bibinfo{author}{Simoniello, R.},
  \bibinfo{author}{Knudsen, V.~F.}, \bibinfo{author}{Karoff, C.},
  \bibinfo{author}{Olsen, J.}, \bibinfo{author}{Turck-Chi\'eze, S.}, \&
  \bibinfo{author}{Jacobsen, B.~H.} (\bibinfo{year}{2015}).
\newblock \bibinfo{title}{Grand solar minima and maxima deduced from $^{10}$be
  and $^{14}$c: magnetic dynamo configuration and polarity reversal}.
\newblock {\it \bibinfo{journal}{Astron. Astrophys.}\/},  {\it
  \bibinfo{volume}{577}\/}, \bibinfo{pages}{{A20}}.
\bibitem[{Kremliovsky(1995)}]{kremliovsky95}
\bibinfo{author}{Kremliovsky, M.} (\bibinfo{year}{1995}).
\newblock \bibinfo{title}{Limits of predictability of solar activity}.
\newblock {\it \bibinfo{journal}{Solar Phys.}\/},  {\it
  \bibinfo{volume}{159}\/}, \bibinfo{pages}{371--380}.
  \DOIprefix\doi{10.1007/BF00686538}.
\bibitem[{{Muscheler} et~al.(2016){Muscheler}, {Adolphi}, {Herbst} \&
  {Nilsson}}]{muscheler16}
\bibinfo{author}{{Muscheler}, R.}, \bibinfo{author}{{Adolphi}, F.},
  \bibinfo{author}{{Herbst}, K.}, \& \bibinfo{author}{{Nilsson}, A.}
  (\bibinfo{year}{2016}).
\newblock \bibinfo{title}{{The Revised Sunspot Record in Comparison to
  Cosmogenic Radionuclide-Based Solar Activity Reconstructions}}.
\newblock {\it \bibinfo{journal}{Solar Phys.}\/},  {\it
  \bibinfo{volume}{291}\/}, \bibinfo{pages}{3025--3043}.
  \DOIprefix\doi{10.1007/s11207-016-0969-z}.
\bibitem[{Ogurtsov et~al.(2002)Ogurtsov, Nagovitsyn, Kocharov \&
  Jungner}]{ogurtsov02}
\bibinfo{author}{Ogurtsov, M.}, \bibinfo{author}{Nagovitsyn, Y.},
  \bibinfo{author}{Kocharov, G.}, \& \bibinfo{author}{Jungner, H.}
  (\bibinfo{year}{2002}).
\newblock \bibinfo{title}{Long-period cycles of the sun's activity recorded in
  direct solar data and proxies}.
\newblock {\it \bibinfo{journal}{Solar Phys.}\/},  {\it
  \bibinfo{volume}{211}\/}, \bibinfo{pages}{371--394}.
\bibitem[{{Petrovay}(2010)}]{petrovay10}
\bibinfo{author}{{Petrovay}, K.} (\bibinfo{year}{2010}).
\newblock \bibinfo{title}{{Solar Cycle Prediction}}.
\newblock {\it \bibinfo{journal}{Living Rev. Solar Phys.}\/},  {\it
  \bibinfo{volume}{7}\/}, \bibinfo{pages}{6}.
  \href{http://arxiv.org/abs/1012.5513}{\tt arXiv:1012.5513}.
\bibitem[{{Popova} et~al.(2017){Popova}, {Zharkova}, {Shepherd} \&
  {Zharkov}}]{popova17}
\bibinfo{author}{{Popova}, E.}, \bibinfo{author}{{Zharkova}, V.},
  \bibinfo{author}{{Shepherd}, S.}, \& \bibinfo{author}{{Zharkov}, S.}
  (\bibinfo{year}{2017}).
\newblock \bibinfo{title}{On a role of quadruple component of magnetic field in
  defining solar activity in grand cycles}.
\newblock {\it \bibinfo{journal}{J. Atmosph. Solar-Terr. Phys.}\/},  (p.
  \bibinfo{pages}{{(in press)}}). \DOIprefix\doi{j.jastp.2017.05.006}.
\bibitem[{Schove(1955)}]{schove55}
\bibinfo{author}{Schove, D.} (\bibinfo{year}{1955}).
\newblock \bibinfo{title}{The sunspot cycle, 649 b.c. to a.d. 2000}.
\newblock {\it \bibinfo{journal}{J.Geophys.Res.}\/},  {\it
  \bibinfo{volume}{60}\/}, \bibinfo{pages}{127--146}.
\bibitem[{Steinhilber et~al.(2012)Steinhilber, Abreu, Beer, Brunner, Christl,
  Fischer, Heikkilae, Kubik, Mann, McCracken, Miller, Miyahara, Oerter \&
  Wilhelms}]{steinhilber12}
\bibinfo{author}{Steinhilber, F.}, \bibinfo{author}{Abreu, J.},
  \bibinfo{author}{Beer, J.}, \bibinfo{author}{Brunner, I.},
  \bibinfo{author}{Christl, M.}, \bibinfo{author}{Fischer, H.},
  \bibinfo{author}{Heikkilae, U.}, \bibinfo{author}{Kubik, P.},
  \bibinfo{author}{Mann, M.}, \bibinfo{author}{McCracken, K.},
  \bibinfo{author}{Miller, H.}, \bibinfo{author}{Miyahara, H.},
  \bibinfo{author}{Oerter, H.}, \& \bibinfo{author}{Wilhelms, F.}
  (\bibinfo{year}{2012}).
\newblock \bibinfo{title}{9,400 years of cosmic radiation and solar activity
  from ice cores and tree rings}.
\newblock {\it \bibinfo{journal}{Proc. Nat. Acad. Sci. USA}\/},  {\it
  \bibinfo{volume}{109}\/}, \bibinfo{pages}{5967--5971}.
  \DOIprefix\doi{10.1073/pnas.1118965109}.
\bibitem[{{Usoskin}(2017)}]{usoskin_LR_17}
\bibinfo{author}{{Usoskin}, I.~G.} (\bibinfo{year}{2017}).
\newblock \bibinfo{title}{{A History of Solar Activity over Millennia}}.
\newblock {\it \bibinfo{journal}{Living Rev. Solar Phys.}\/},  {\it
  \bibinfo{volume}{14}\/}, \bibinfo{pages}{3}.
  \DOIprefix\doi{10.1007/s41116-017-0006-9}.
\bibitem[{{Usoskin} et~al.(2014){Usoskin}, {Hulot}, {Gallet}, {Roth}, {Licht},
  {Joos}, {Kovaltsov}, {Th{\'e}bault} \& {Khokhlov}}]{usoskin_AAL_14}
\bibinfo{author}{{Usoskin}, I.~G.}, \bibinfo{author}{{Hulot}, G.},
  \bibinfo{author}{{Gallet}, Y.}, \bibinfo{author}{{Roth}, R.},
  \bibinfo{author}{{Licht}, A.}, \bibinfo{author}{{Joos}, F.},
  \bibinfo{author}{{Kovaltsov}, G.~A.}, \bibinfo{author}{{Th{\'e}bault}, E.},
  \& \bibinfo{author}{{Khokhlov}, A.} (\bibinfo{year}{2014}).
\newblock \bibinfo{title}{{Evidence for distinct modes of solar activity}}.
\newblock {\it \bibinfo{journal}{Astron. Astrophys.}\/},  {\it
  \bibinfo{volume}{562}\/}, \bibinfo{pages}{{L10}}.
  \DOIprefix\doi{10.1051/0004-6361/201423391}.
  \href{http://arxiv.org/abs/1402.4720}{\tt arXiv:1402.4720}.
\bibitem[{{Usoskin} et~al.(2006){Usoskin}, {Voiculescu}, {Kovaltsov} \&
  {Mursula}}]{usoskin_JASTP_06}
\bibinfo{author}{{Usoskin}, I.~G.}, \bibinfo{author}{{Voiculescu}, M.},
  \bibinfo{author}{{Kovaltsov}, G.~A.}, \& \bibinfo{author}{{Mursula}, K.}
  (\bibinfo{year}{2006}).
\newblock \bibinfo{title}{{Correlation between clouds at different altitudes
  and solar activity: Fact or Artifact?}}
\newblock {\it \bibinfo{journal}{J. Atmosph. Sol.-Terr. Phys.}\/},  {\it
  \bibinfo{volume}{68}\/}, \bibinfo{pages}{2164--2172}.
  \DOIprefix\doi{10.1016/j.jastp.2006.08.005}.
\bibitem[{{Vecchio} et~al.(2017){Vecchio}, {Lepreti}, {Laurenza}, {Alberti} \&
  {Carbone}}]{vecchio17}
\bibinfo{author}{{Vecchio}, A.}, \bibinfo{author}{{Lepreti}, F.},
  \bibinfo{author}{{Laurenza}, M.}, \bibinfo{author}{{Alberti}, T.}, \&
  \bibinfo{author}{{Carbone}, V.} (\bibinfo{year}{2017}).
\newblock \bibinfo{title}{{Connection between solar activity cycles and grand
  minima generation}}.
\newblock {\it \bibinfo{journal}{Astron. Astrophys}\/},  {\it
  \bibinfo{volume}{599}\/}, \bibinfo{pages}{{A58}}.
  \DOIprefix\doi{10.1051/0004-6361/201629758}.
\bibitem[{Zharkova et~al.({2015})Zharkova, Shepherd, Popova \&
  Zharkov}]{zharkova15}
\bibinfo{author}{Zharkova, V.~V.}, \bibinfo{author}{Shepherd, S.~J.},
  \bibinfo{author}{Popova, E.}, \& \bibinfo{author}{Zharkov, S.~I.}
  (\bibinfo{year}{{2015}}).
\newblock \bibinfo{title}{{Heartbeat of the Sun from Principal Component
  Analysis and prediction of solar activity on a millenium timescale}}.
\newblock {\it \bibinfo{journal}{{Sci. Rep.}}\/},  {\it
  \bibinfo{volume}{{5}}\/}, \bibinfo{pages}{{15689}}.
  \DOIprefix\doi{{10.1038/srep15689}}.
\bibitem[{{Zharkova} et~al.(2017){Zharkova}, {Shepherd}, {Popova} \&
  {Zharkov}}]{zharkova17}
\bibinfo{author}{{Zharkova}, V.~V.}, \bibinfo{author}{{Shepherd}, S.~J.},
  \bibinfo{author}{{Popova}, E.}, \& \bibinfo{author}{{Zharkov}, S.~I.}
  (\bibinfo{year}{2017}).
\newblock \bibinfo{title}{{Reinforcing the double dynamo model with
  solar-terrestrial activity in the past three millennia}}.
\newblock {\it \bibinfo{journal}{ArXiv e-prints}\/}, .
  \href{http://arxiv.org/abs/1705.04482}{\tt arXiv:1705.04482}.

\end{thebibliography}

\end{document}